\documentclass[prb,amsmath,twocolumn,amssymb,superscriptaddress]{revtex4}

\usepackage{amsmath}
\usepackage{graphicx}
\usepackage{multirow}
\usepackage{subfig}
\usepackage{bbm,color,ulem}
\usepackage{dsfont}
\usepackage{float}
\usepackage{braket}
\usepackage[export]{adjustbox}
\allowdisplaybreaks

\newcommand{\ua}{\uparrow}
\newcommand{\da}{\downarrow}

\DeclareMathOperator{\tr}{Tr}

\newcommand{\numb}{\addtocounter{equation}{1}\tag{\theequation}}

\begin{document}

\title{Magnetic Phases of Bilayer Quantum-Dot Hubbard Model Plaquettes}

\author{Donovan Buterakos}
\author{Sankar Das Sarma}
\affiliation{Condensed Matter Theory Center and Joint Quantum Institute, Department of Physics, University of Maryland, College Park, Maryland 20742-4111 USA}
\date{\today}

\begin{abstract}

It has been demonstrated that small plaquettes of quantum dot spin qubits are capable of simulating condensed matter phenomena which arise from the Hubbard model, such as the collective Coulomb blockade and Nagaoka ferromagnetism. Motivated by recent materials developments, we investigate a bilayer arrangement of quantum dots with four dots in each layer which exhibits a complex ground state behavior. We find using a generalized Hubbard model with long-range Coulomb interactions, several distinct magnetic phases occur as the Coulomb interaction strength is varied, with possible ground states that are ferromagnetic, antiferromagnetic, or having both one antiferromagnetic and one ferromagnetic layer. We map out the full phase diagram of the system as it depends on the inter- and intra-layer Coulomb interaction strengths, and find that for a single layer, a similar but simpler effect occurs. We also predict interesting contrasts among electron, hole, and electron-hole bilayer systems arising from complex correlation physics. Observing the predicted magnetic configuration in already-existing few-dot semiconductor bilayer structures could prove to be an important assessment of current experimental quantum dot devices, particularly in the context of spin-qubit-based analog quantum simulations.

\end{abstract}

\maketitle

\section{Introduction}

Quantum dots remain a promising platform for developing quantum information technologies due to their long coherence times (on the order of seconds in Si devices) \cite{MiSCI2017, ChatterjeeNRP2021} and their small physical size which allows for long-term device scalability. These advantages are among the key criteria for the implementation of a quantum processor \cite{DiVincenzoFP2000}. Since their initial proposal as a quantum computing platform in 1998 \cite{LossPRA1998}, there has been slow but steady progress in the fabrication and control of quantum devices as many obstacles have been overcome\cite{BurkardRMP2023}. Over the past few years, there has been progress in fabricating quantum dot arrays in Silicon \cite{UnseldARXIV2023} and Germanium \cite{LawrieAPL2020, vanRiggelenAPL2021}. A 3x3 Phosphorus dopant array has been operated \cite{WangNC2022}, and a 4x4 quantum dot array has been fabricated in Germanium \cite{BorsoiARXIV2022}. Full control has been demonstrated over a six-qubit quantum processor \cite{PhilipsNAT2022}. Quantum dot based spin qubits and quantum processors are among the three most active and important current quantum computing platforms (along with superconducting transmons and atomic ion traps).

Despite the successes of quantum dot devices, experimental capabilities are still well below what is needed for commercial applications such as Shor's algorithm \cite{ShorIEEE1994, ShorSIAMJC1997}. Quantum error correction protocols such as surface code \cite{KitaevAP2003, RaussendorfNJP2007, FowlerPRA2009, WangQIC2010} allow for the possibility of fault-tolerant quantum computing, but necessitate error thresholds well below 1\% and require thousands of physical qubits for each logical qubit. Current devices still struggle with the presence of charge noise \cite{DuttaRMP1981, HuPRL2006, HuPRB2011, HollmannPRApp2020} as well as challenges tuning qubits in larger devices, which will need to be fully automated in order to operate devices suitable for industrial applications \cite{BotzemPRApp2018, vanDiepenAPL2018, ZwolakPRApp2020}. Because the fault-tolerant implementation of quantum algorithms is still many years (if not decades) away, it is crucial to discuss the applications of current qubit technologies, small plaquettes of around 4-9 quantum dots, to interesting quantum problems.

One of the successful models to theoretically investigate realistic quantum dot qubit systems is the generalized Hubbard model \cite{HubbardPRS1963, ShuoPRB2011, DasSarmaPRB2011, WangPRB2011}. The Hubbard Model is a minimal model that was originally proposed to study ferromagnetism in transition metals and excels in explaining a host of other condensed-matter phenomena such as antiferromagnetism and Mott transitions. However, a full understanding of ferromagnetism starting from a microscopic model still eludes us except in a few special cases (in fact, the Hubbard model generically gives rise to antiferromagnetism rather than ferromagnetism). These ferromagnetic cases include Nagaoka ferromagnetism, which is a mathematical theorem that predicts ground state ferromagnetism for systems precisely 1 electron above half-filling for certain graphs of lattice sites \cite{NagaokaPR1966, TasakiPTP1998}, and flatband ferromagnetism, which is a similar result for bands which have a large number of degenerate states \cite{TasakiPTP1998, MielkeJPA1991, MielkeJPA1992, TasakiPRL1992}. These are both of course rather idealized (and unrealistic) situations for macroscopic systems. It was first proposed in 1994 that the Hubbard model could be simulated by a quantum plaquette \cite{StaffordPRL1994}, and was later realized experimentally when the collective Coulomb blockade was observed in a Hubbard model simulated by a quantum dot array \cite{HensgensNAT2017}. In 2020, Nagaoka ferromagnetism was also observed in a 2x2 quantum dot array \cite{DehollainNAT2020}. Observing physical phenomena such as these in models simulated by quantum dot plaquettes can act as intermediate milestones between single-qubit devices and a full fault-tolerant quantum processor capable of running quantum algorithms. As an aside we mention that there has been enormous recent progress in the simulation of the fermionic Hubbard model in cold atomic systems, where the physics and the experimental constraints are completely different from the quantum dot systems of our interest in the current work \cite{MazurenkoNAT2017, SompetNAT2022}. Hubbard model ferromagnetism has, however, not been studied in the cold fermionic atomic systems.

In a recent breakthrough work at Delft, operation of a vertical gate-defined double quantum dot was demonstrated in a Ge/SiGe double quantum well \cite{TidjaniARXIV2023}. This new result opens the possibility of creating quantum dot devices extending beyond a single plane, and beyond just electrons since Ge introduces holes. By combining several vertical double quantum dots in a 2-dimensional array, the authors of Ref. \onlinecite{TidjaniARXIV2023} envision that a full bilayer device can be made. Such a device could have many potential applications ranging from improving quantum computing architectures to simulating bilayer many-body physics.

In this paper, motivated by the Delft experiment, we investigate the magnetic properties of the Hubbard model ground states of a bilayer quantum dot system. We give a specific example of a bilayer plaquette where each layer contains 4 quantum dots arranged along a triangular lattice (other configurations can also be studied quite generally using the same technique). We show that if a bilayer quantum dot device were to simulate a Hubbard model with such a geometry (similarly to in Ref. \onlinecite{DehollainNAT2020}), then nontrivial and unexpected magnetic behavior of the ground state would arise from the inter-dot correlations. Using exact diagonalization, we determine the spin configuration in each layer as a function of the inter- and intralayer Coulomb interaction strengths $V$ and $V'$. We find that magnetic phase transitions occur in both variables and that the system behaves much differently depending on the relative strengths of the interactions. Of course, these are only finite size magnetic phases and transitions among them, perhaps more akin to magnetic phases of various molecules occurring in this bilayer dot system. We hope that our work will motivate future bilayer quantum dot experiments, as we predict that highly nontrivial magnetic behavior should be observable with a plaquette of only 4 vertical double quantum dots, which is likely within the scope of experimental capabilities over the next few years, if not already right now.  Our main motivation is to introduce nontrivial theoretical ideas which can be applied to the existing small system quantum dot circuits to do useful quantum many-body physics as is being done extensively using both ion trap and superconducting qubits.

This paper is organized as follows: in Sec. II, we introduce and define the relevant Hubbard model Hamiltonian for our system. In Sec. III, we give our results, starting with special cases where $V$ or $V'$ equals $0$, then showing the complete phase diagram with $V$ and $V'$ both variable, and lastly investigating a similar system filled with electrons rather than holes. Finally, we give our conclusions in Sec. IV.

\begin{figure}[!htb]
	\includegraphics[width=.99\columnwidth]{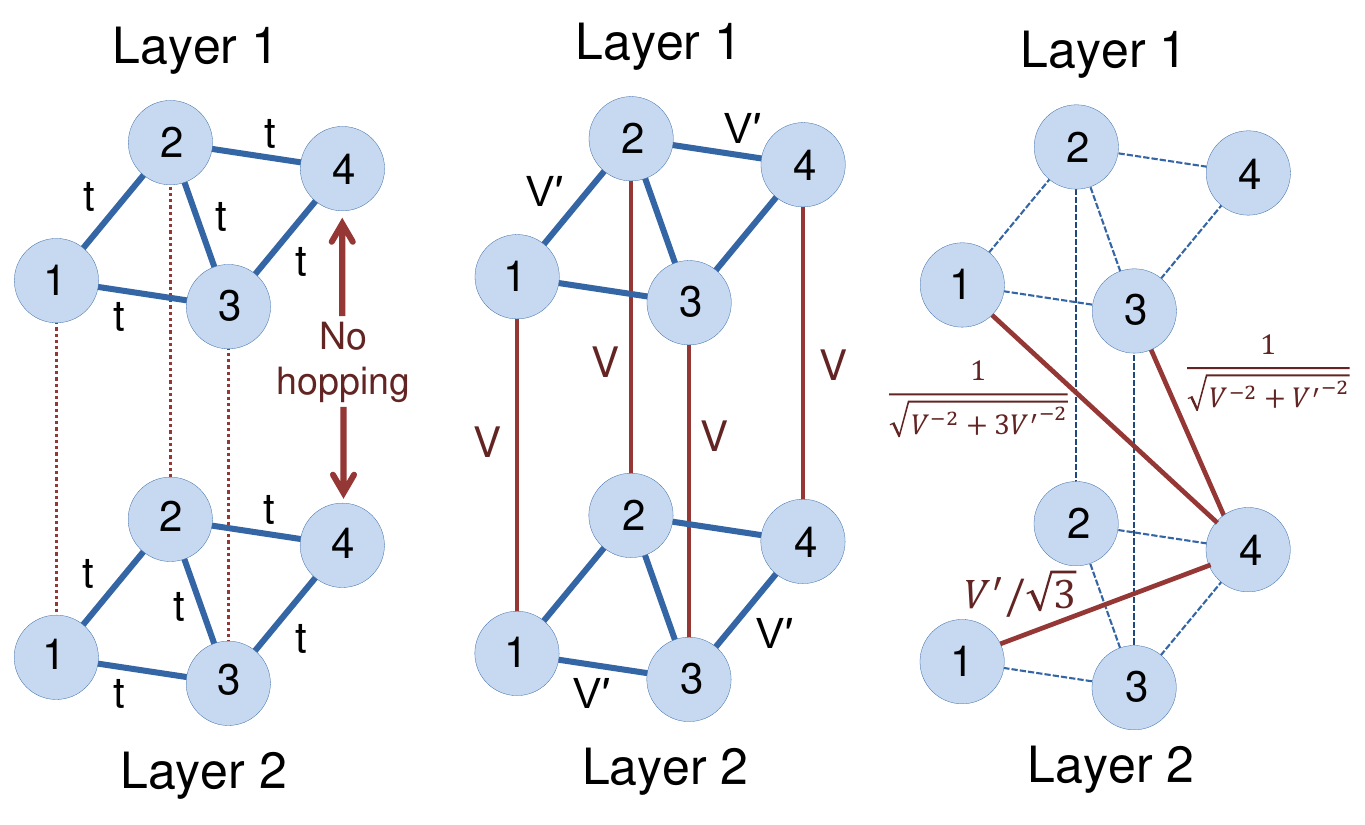}
	\caption{{\bf Left:} Bilayer quantum dot plaquette with a rhombus geometry. There is a tunneling coefficient $t$ between adjacent dots within a single layer, but no tunneling between layers. {\bf Center:} Interlayer Coulomb interactions with strength $V$ and intralayer Coulomb interactions with strength $V'$. {\bf Right:} Long-range Coulomb interactions are present with strengths written in terms of $V$ and $V'$, see eq. (\ref{eqn:v}).}
	\label{fig:layout}
\end{figure}

\section{Model and Hamiltonian}

We consider a bilayer plaquette layout where each layer contains four quantum dots arranged in a rhombus, with dots from layer 1 being positioned directly above dots in layer 2, as depicted in fig. \ref{fig:layout}. We allow tunneling between adjacent dots within a single layer, but disallow tunneling between layers, consistent with the Delft experimental system. We include an onsite Coulomb interaction term $U$, as well as long-range Coulomb interaction terms between all dots in both layers. This system can be modeled with the following generalized Hubbard Hamiltonian, where $c_{l,i\sigma}^{\dagger}$ is the creation operator for a particle / hole on site $i$ of layer $l$ with spin $\sigma$, and $n_{l,i\sigma}=c_{l,i\sigma}^{\dagger}c_{l,i\sigma}$:

\begin{align*}
H&=\sum_{\substack{i\ne j\\l,\sigma}} t_{l,ij}c_{l,i\sigma}^{\dagger} c_{l,j\sigma} + U \sum_{l,i}n_{l,i\ua}n_{l,i\da} \\ &+ \!\!\!\!\!\!\sum_{\substack{(l,i)\ne (m,j)\\\sigma,\tau}}\!\!\!\!\!\!\frac{V_{lm,ij}}{2}n_{l,i\sigma}n_{m,j\tau}
\numb
\label{eqn:h}
\end{align*}

Because there is no tunneling between layers, total particle number within a single layer is conserved. We consider the cases where each layer contains either two holes or two electrons. We note that for the single-band Hubbard model, having two holes is equivalent to being filled with six electrons and vice versa since there are 8 states for 4 spinful fermions in our plaquette. For the sake of simplicity, we will assume the tunneling coefficients between all adjacent pairs of dots have the same amplitude $t$. This can be relaxed at the considerable cost of the final results becoming much more complicated than they already are.  In any case, variable t values would correspond to a random incoherent system, which is not a particularly meaningful system to study. The sign of $t_{l,ij}$ will depend on whether the layer contains holes or electrons, as follows:

\begin{equation}
t_{l,ij}=\begin{cases}
t&\text{for adjacent $i,j$ if layer $l$ contains holes}\\
-t&\text{for adjacent $i,j$ if layer $l$ contains electrons}\\
0&\text{for nonadjacent $i,j$}
\end{cases}
\label{eqn:t}
\end{equation}

The separation between layers can be controlled when fabricating devices, and thus the interlayer Coulomb interaction strength can vary compared to the intralayer interaction strength. Note that, by contrast, the intralayer interaction (as well as the hopping $t$) depends on the dot placement in the layer. We define $V$ to be the nearest-neighbor Coulomb interaction strength between layers, and we let $V'$ be the interaction strength between neighboring dots within a single layer. The Coulomb interaction strength over longer distances is given in terms of these two scales according to a simple $1/r$ potential. Thus $V_{lm,ij}$ for the rhombus plaquette is given as follows:

\begin{equation}
V_{lm,ij}=\begin{cases}
V'&\text{for $l=m$ and adjacent $i,j$}\\
V'/\sqrt{3}&\text{for $l=m$ and nonadjacent $i,j$}\\
\pm V&\text{for $l\ne m$ and $i=j$}\\
\frac{\pm 1}{\sqrt{V^{-2}+V'^{-2}}}&\text{for $l\ne m$ and adjacent $i,j$}\\
\frac{\pm 1}{\sqrt{V^{-2}+3V'^{-2}}}&\text{for $l\ne m$ and nonadjacent $i,j$}
\end{cases}
\label{eqn:v}
\end{equation}

where the $\pm$ sign is determined by the charge of the particles (i.e. electron/hole) in each layer. We will assume that the onsite Coulomb interaction strength is much larger than the other energy scales in the model, that is $U \gg V,V',t$. For this work we neglect the higher-energy states with doubly-occupied dots, as these states contribute corrections on the order of the exchange interaction $O(t^2/U)$, which are much smaller than the other energy scales present in the system (particularly since $t$ tends to be small in quantum dot qubits). We have investigated the effects of these higher-energy states in previous works \cite{ButerakosPRB2019, ButerakosPRB2023}, and found that the magnetic behavior of the system remains unchanged so long as $U/t$ is larger than some geometry-dependent critical value. Note that while the absence of doubly-occupied dots makes the Hubbard model with $U=\infty$ at half filling equivalent to a Heisenburg model, away from half-filling the Hubbard model still cannot be represented by a Heisenburg model, since the Heisenburg model does not permit dots to be empty. Unoccupied dots are crucial to magnetic effects such as Nagaoka Ferromagnetism, which comes about due to the kinetic motion of electrons or holes within the plaquette \cite{TasakiPTP1998}. Thus, the physics of interest here is much richer than the usual exchange-coupled spin qubits considered in most quantum dot qubit architecture as our goal is to use the quantum dot platform to study complex quantum magnetism by going beyond the strict spin-spin coupled Heisenberg model.

\section {Results}

We first examine the case where both layers are occupied by holes, and hence $t_{l,ij}>0$ and the Coulomb interaction between layers is repulsive ($V_{lm,ij}>0$). Because there is no tunneling between layers (this is the experimental situation) and hence no exchange interaction between layers, spin is conserved within each layer. Therefore all eigenstates have one of the following spin configurations: both layers have total spin 0, both layers have total spin 1, or one layer is spin 0 while the other is spin 1. In the latter case, there will be a trivial two-fold degeneracy since in our model the two layers are identical--that is, any eigenstate with spin configuration $(0,1)$ would have the same energy as $(1,0)$. In practice, this degeneracy may be broken due to asymmetries in the physical system, however for the sake of this discussion, we will consider only one of these spin configurations, as the physics for the other are similar. The situation with strong asymmetries is theoretically uninteresting since it is basically a random system with the results varying from sample to sample with no generic ground states to discuss.

We use exact diagonalization to determine the ground state of the Hamiltonian for various choices of the parameters $V$ and $V'$, while keeping $t$ fixed. We specifically note the total spin of each layer, with particular interest in the question of how the ground state spin relies on the parameters $V/t$ and $V'/t$. We first investigate these parameters individually, and then give a phase diagram of the system while allowing both parameters to vary independently. Thus, the dimensionless intralayer and interlayer Coulomb couplings determine the quantum phases of the system.

\subsection{$V'=0$ Limit}

\begin{figure}[!htb]
	\includegraphics[width=.99\columnwidth]{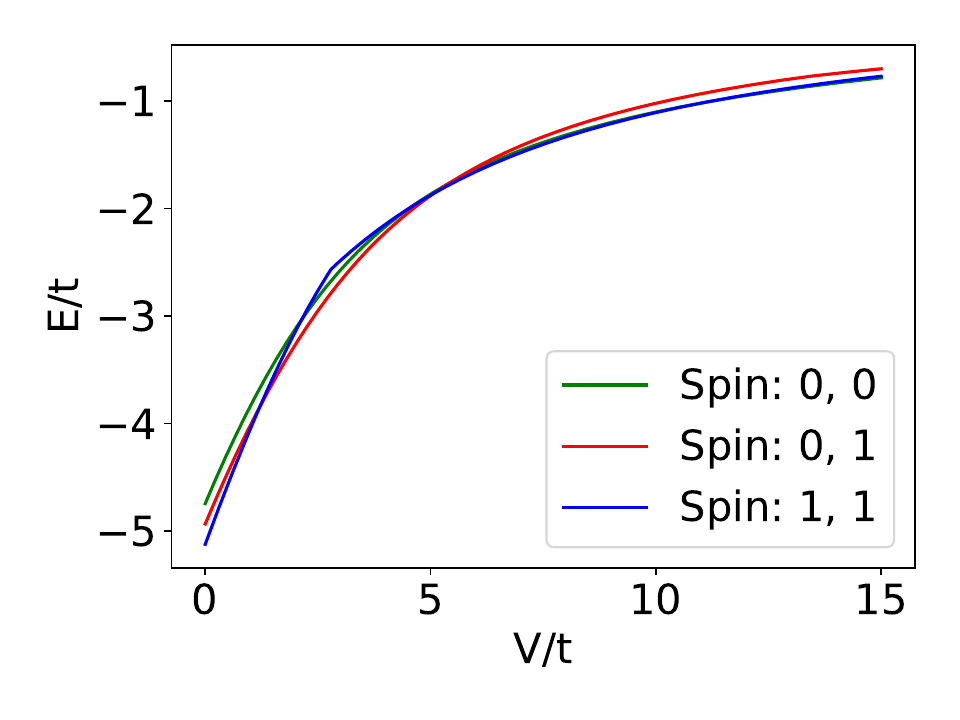}
	\includegraphics[width=.49\columnwidth]{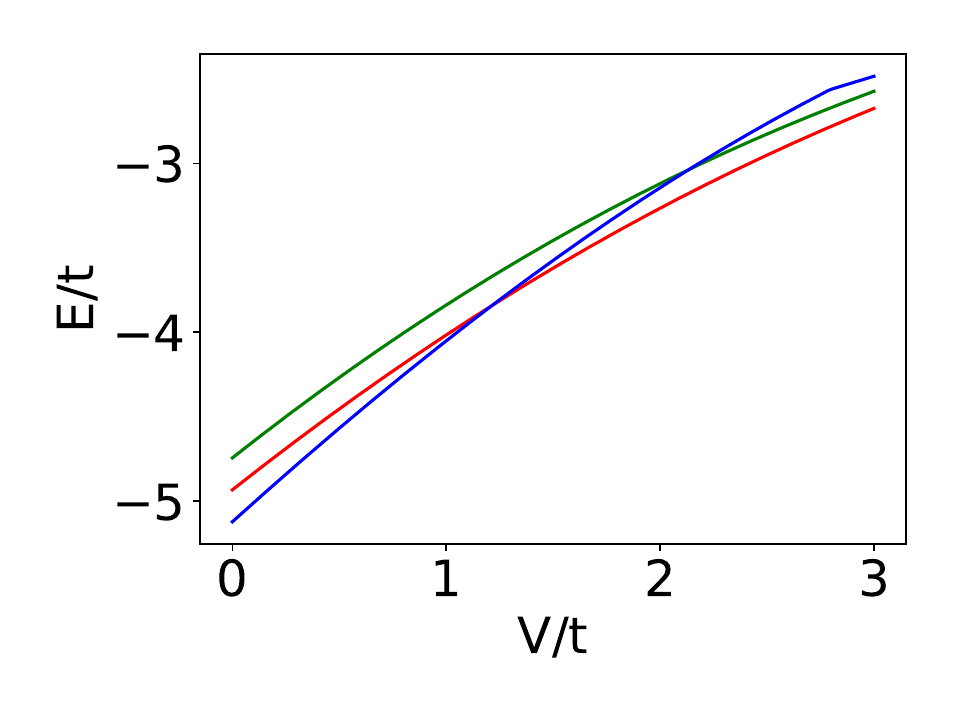}
	\includegraphics[width=.49\columnwidth]{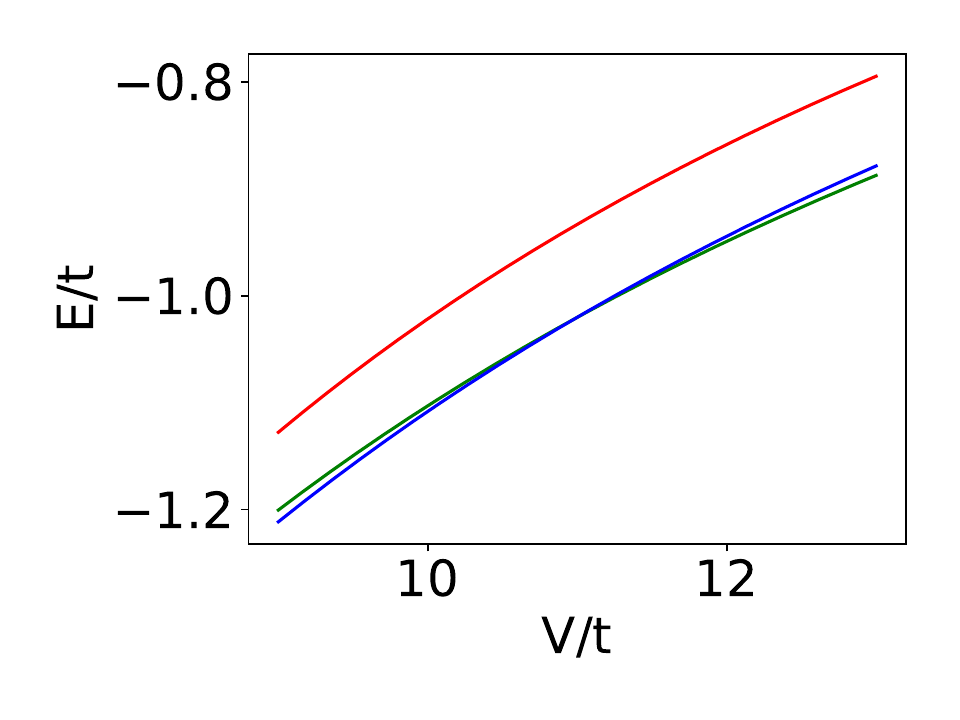}
	\caption{{\bf Top:} Ground state energy versus $V/t$ for each of the three possible spin-configurations. {\bf Bottom 2:} Zoom-in on various crossings near $V/t=1.2$ and $V/t=11$.}
	\label{fig:vp0spin}
\end{figure}

We first consider the limit where $V'=0$. This represents a system where the separation between layers is significantly smaller than the distance between dots within a layer. In fig. \ref{fig:vp0spin}, we plot the lowest energy eigenstate in each of the three spin configurations versus $V$, holding $t$ constant. Remarkably, the ground state changes spin configuration several times as $V$ changes. Initially, when $V=0$, the system behaves like 2 completely noninteracting layers, each of which having a spin 1 ground state. However, at $V=1.2t$, the system transitions to the spin $(0,1)$ configuration. At $V=5.1t$, the system transitions back to a spin $(1,1)$ ground state, although as we will discuss, this second spin $(1,1)$ ground state is qualitatively different from the initial spin $(1,1)$ ground state. Finally, at $V=11t$ and onwards, the system adopts a completely antiferromagnetic ground state in the strong interaction limit (as is expected in a Hubbard model).

\begin{figure}[!htb]
	\includegraphics[width=.99\columnwidth]{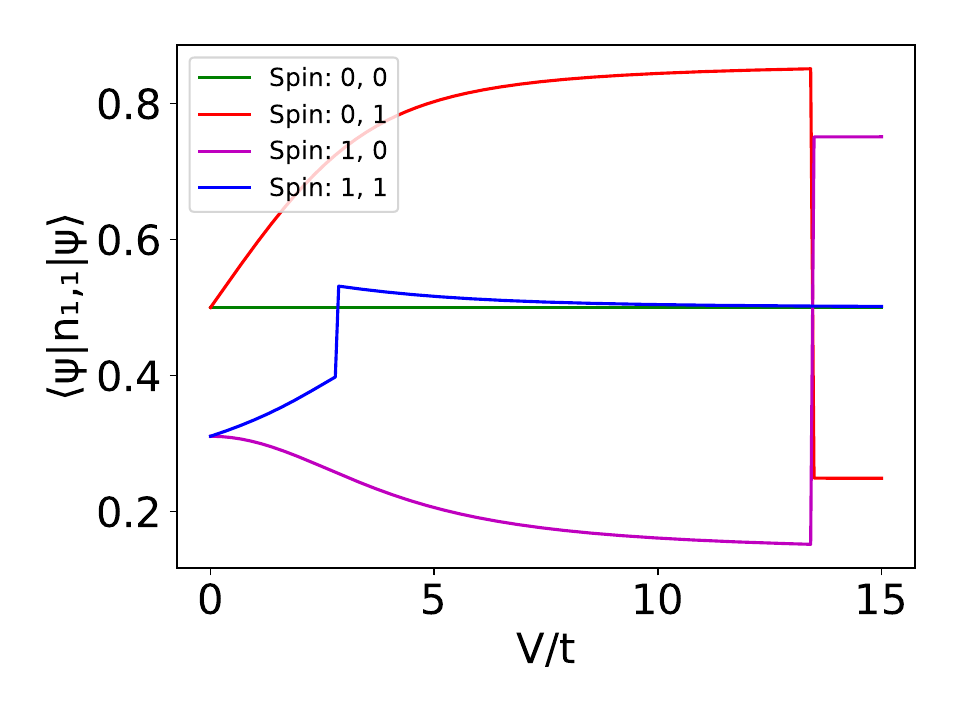}
	\caption{$\bra{\Psi}n_{1,1}\ket{\Psi}$ for dot 1 of layer 1 (one of the outer dots), for each possible spin configuration. Note that the degenerate (0, 1) and (1, 0) spin configurations are plotted separately here, since the locations of the holes in each layer are different.}
	\label{fig:vp0n}
\end{figure}

In order to understand the transition between the $(1,1)$ and $(0,1)$ spin configurations, it is helpful to examine the orbital dependency of the wavefunctions. In fig. \ref{fig:vp0n}, we plot the expectation value of the occupation number of dot 1 (one of the outer dots) $\bra{\Psi}n_{l,1}\ket{\Psi}$, where $\ket{\Psi}$ is the ground state wavefunction for a particular spin configuration. Note that due to symmetry each layer will behave identically, and within each layer, the outer dots (1 and 4) will have the same average occupation number, and the inner dots (2 and 3) will have the same occupation number. Additionally, since total particle number is conserved within layers, for 2 holes the expectation value of all four dots must add up to 2, and therefore $\langle n_{l,1} \rangle = \langle n_{l,4} \rangle = 1 - \langle n_{l,2} \rangle = 1 - \langle n_{l,3} \rangle$.

From fig. \ref{fig:vp0n}, we see that in the absence of Coulomb interactions (that is when $V=0$), the lowest-energy spin 0 and spin 1 states in each layer have differing dot occupations. This is a feature of the specific plaquette geometry and particle number (2 holes in a 4-dot rhombus geometry), and varies in general. When the two holes within a layer form a spin singlet, the multiparticle wavefunction distributes the two holes evenly among the four dots, and thus each dot has an average occupation of $1/2$. In contrast, the spin 1 ground state distributes the holes unevenly, with a 0.69 probability of observing the inner dots 2 and 3 occupied, and a 0.31 probability of observing the outer dots 1 and 4 occupied. This imbalance between the ground state occupation of the inner and outer dots, and specifically the fact that this imbalance is dependent on the total spin in each layer, gives rise to interesting physical effects.

The first phase transition at $V=1.2t$ arises from the ground state occupation imbalance. When $V=0$, the two layers are decoupled, and thus each independently adopts a spin 1 ground state. As $V$ is then increased, the layers become weakly coupled, and an additional potential energy is added to each state given by the third term of eq. (\ref{eqn:h}), which is dependent on the occupation number of each dot. If the hole density of at least one layer is evenly distributed among the four dots in that layer, then the total additional energy cost from long-range Coulomb interactions between layers will be given by:

\begin{equation}
\langle\Psi|H_V|\Psi\rangle=\sum_i V n_{1,i} n_{2,i} = \sum_i V n_{1,i} (1/2)=V
\end{equation}

since $\sum_in_{l,i}$ is simply the total number of holes in layer $l$, in this case 2. However, if the hole density of both are concentrated towards the center dots, then the long-range Coulomb energy between layers will be greater. In the case of the values given above, it will be given by:

\begin{equation}
\langle\Psi_{11}|H_V|\Psi_{11}\rangle = 2V(.31)^2+2V(.69)^2=1.14V
\end{equation}

Then the ground state of each layer is spin 1 when $V=0$, but this energy increases more quickly than other competing states as $V$ increases. Thus we expect to see a transition to a different spin state, as shown in fig. \ref{fig:vp0spin}.

Note that in fig. \ref{fig:vp0n}, there is a sharp discontinuity in the spin (1, 1) state located at $V=2.8t$. This corresponds directly to the sharp bend in the energy at the same location in fig. \ref{fig:vp0spin}. This is due to a crossing of two spin (1, 1) states, although in our plots, we only plot the lowest-energy spin (1, 1) state. This crossing allows a new spin (1, 1) state to overtake the current ground state as $V$ is increased, causing a transition back to spin (1, 1). A similar but less consequential crossing occurs in the spin (0, 1) state near $V=13t$.

\begin{figure}[!htb]
	\includegraphics[width=.99\columnwidth]{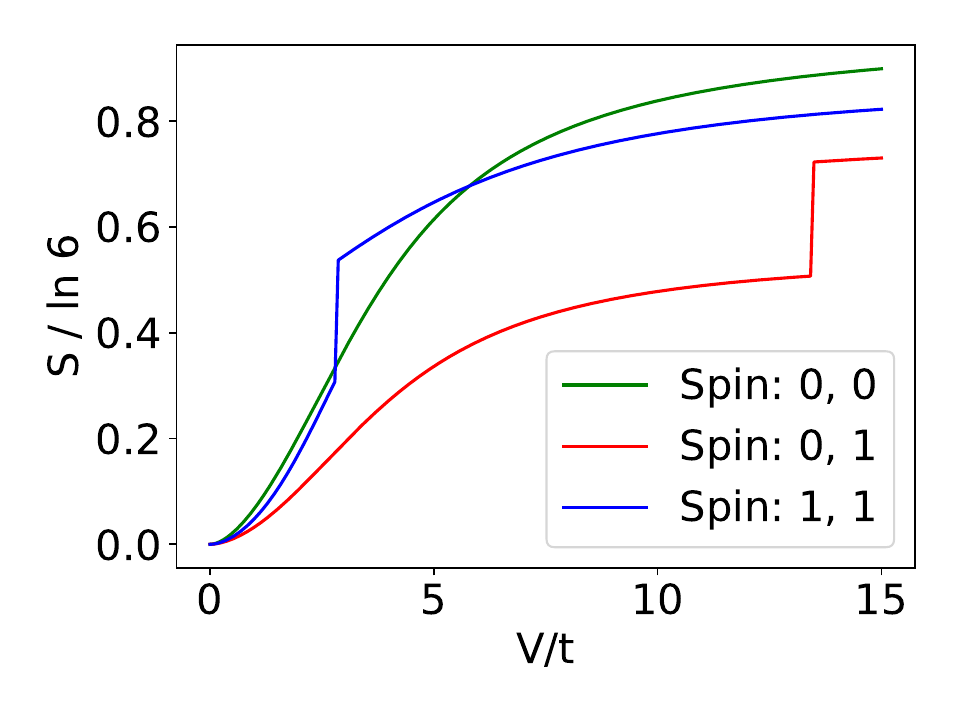}
	\caption{Entanglement entropy between layers, normalized so that a value of 1 indicates maximal entanglement between layers.}
	\label{fig:vp0s}
\end{figure}

It is instructive to consider the Von Neumann entanglement entropy $S$ between layers, defined as follows:

\begin{equation}
S=-\tr_1 \rho_1 \ln \rho_1
\end{equation}

where $\rho_1=\tr_2\ket{\Psi}\bra{\Psi}$, and $\tr_l$ traces out layer $l$. In fig. \ref{fig:vp0s}, we plot the entanglement entropy normalized by the maximum possible entanglement entropy $S_{\text{max}}=\ln 6$, so that a value of 1 indicates maximal entanglement between layers. We find that the entanglement entropy remains small around the first phase transition, which supports the claim that the coupling is weak in this regime. As $V$ increases, the two layers become significantly entangled together and are strongly correlated.

\subsection{$V=0$ Limit}

\begin{figure}[!htb]
	\includegraphics[width=.49\columnwidth]{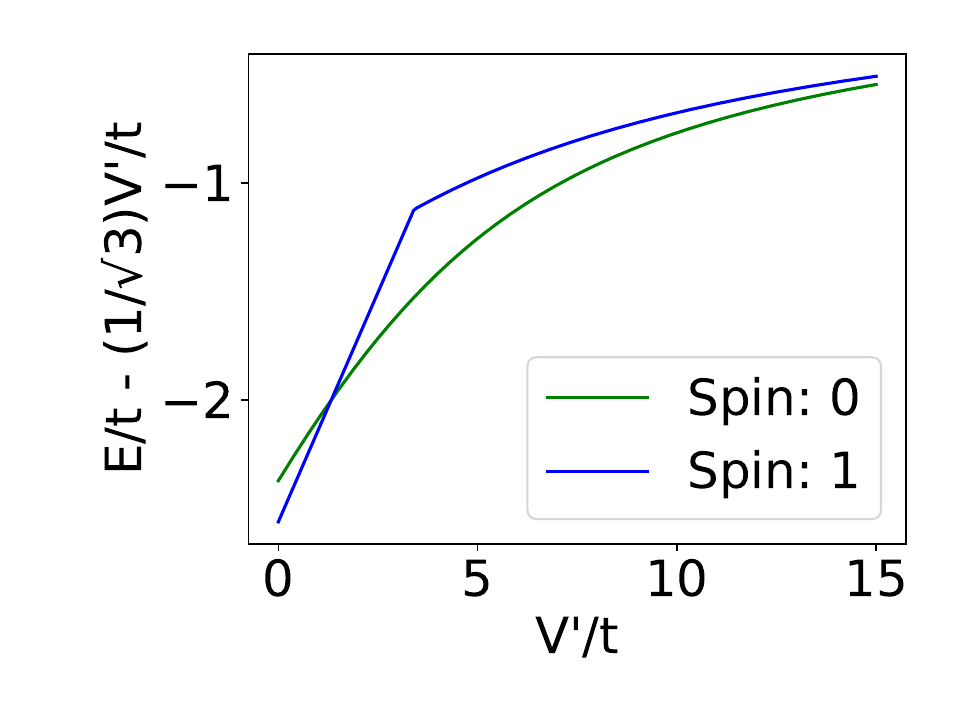}
	\includegraphics[width=.49\columnwidth]{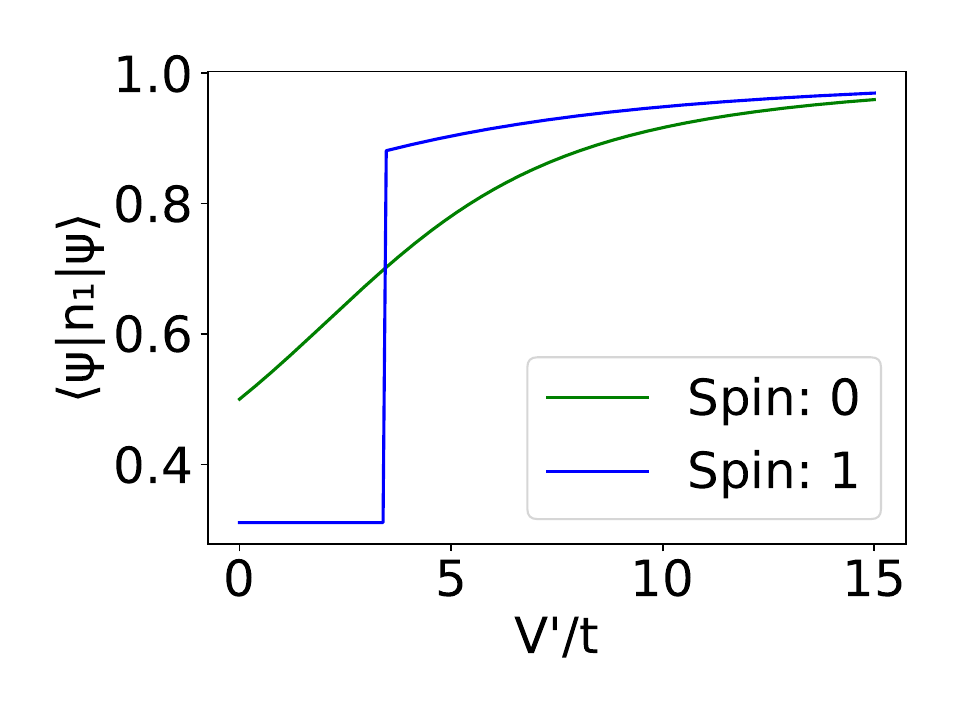}
	\caption{{\bf Left:} Energy for dot 1 of a single layer (one of the outside dots). Note that the trivial linear dependence on the Coulomb interaction strength $\frac{V'}{\sqrt{3}t}$ has been subtracted off. {\bf Right:} Expectation value of the occupation number $\bra{\Psi}n_{1}\ket{\Psi}$.}
	\label{fig:v0en}
\end{figure}

We briefly consider the limit where $V=0$, which is the case when the interlayer separation is large. In this case, the two layers act independently, since our model allows no tunneling between layers. However, a phase transition still exists as $V'$ is changed. In fig. \ref{fig:v0en}, we show the energy and expectation value $\langle n_1\rangle$ of a single rhombus plaquette. Note that the energy has an asymptotically linear (trivial) dependence on $V'$ which has been subtracted off in this figure. There is a single magnetic phase transition from spin 1 to spin 0. This again arises from the the fact that in the absence of long-range Coulomb interactions, the plaquette geometry prefers a spin 1 ground state with holes concentrated in the center two dots of the plaquette (this is just Hund's rule). However, if $V'$ is large, the holes will be repelled away from each other towards the outer two dots of the plaquette. This competition between the Coulomb interaction pushing holes towards the outside, and the plaquette geometry's natural tendency to prefer the holes concentrated towards the center gives rise to a change in spin from 1 to 0 as $V'$ increases, and the system is an antiferromagnet for large enough $V'$.

\subsection{Nonzero $V$ and $V'$}

\begin{figure}[!htb]
	\includegraphics[width=.49\columnwidth]{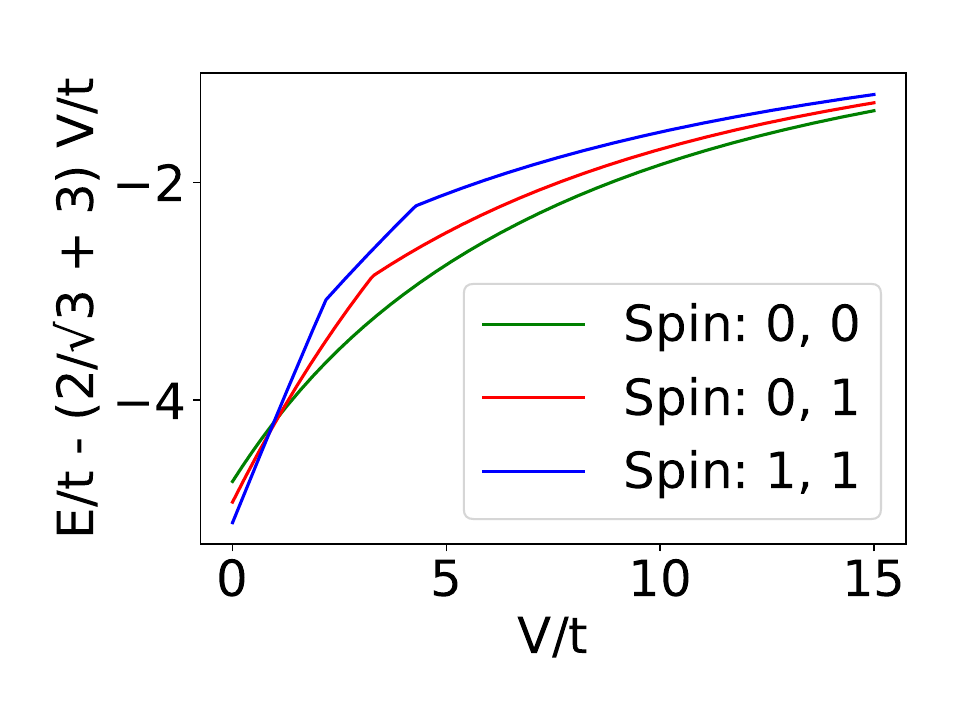}
	\includegraphics[width=.49\columnwidth]{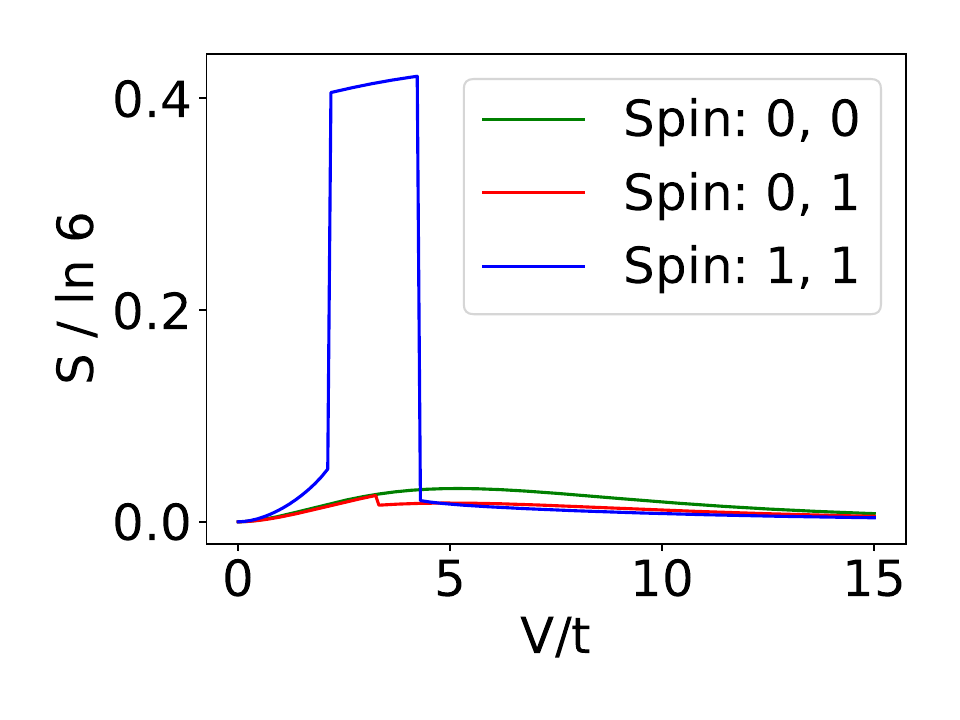}
	\caption{{\bf Left:} Ground state energy for each spin configuration in the case where $V=V'$. Note that the trivial linear dependence on the Coulomb interaction strength $(\frac{2}{\sqrt{3}}+3)\frac{V}{t}$ has been subtracted off. {\bf Right:} Entanglement entropy between layers when $V=V'$, normalized so that a value of 1 indicates maximal entanglement.}
	\label{fig:vequalvp}
\end{figure}

Now consider the case where both the interlayer coupling $V$ and the intralayer coupling $V'$ are nonzero. In fig. \ref{fig:vequalvp}, we plot the ground state energy for each spin configuration in the case where $V=V'$. We find only a single phase transition from the (1, 1) spin configuration to the (0, 0) spin configuration at $V=t$, although the (0, 1) spin configuration briefly becomes the ground state during the transition from (0, 0) to (1, 1). This behavior is similar to two copies of the single-layer phase transition discussed in the previous section. This is made particularly clear by plotting the entanglement entropy, which we also show in fig. \ref{fig:vequalvp}, again normalized so that a value of 1 indicates maximal entanglement between layers. Note that the entanglement entropy stays well below 5\% of its maximal value for all 3 spin configurations except for a brief region between $2t$ and $4t$ where the (1, 1) spin configuration becomes entangled. However, in this region, the (1, 1) spin configuration has a significantly higher energy than the ground state, and thus the entanglement entropy of the ground state remains well below 5\% of maximally-entangled for all values of $V$. This also indicates that for $V=V'$, the system acts as two weakly coupled copies of the single-layer physics.

\begin{figure}[!htb]
	\adjustbox{trim={.05\width} {.1\height} {0} {.2\height},clip}{\includegraphics[width=.99\columnwidth]{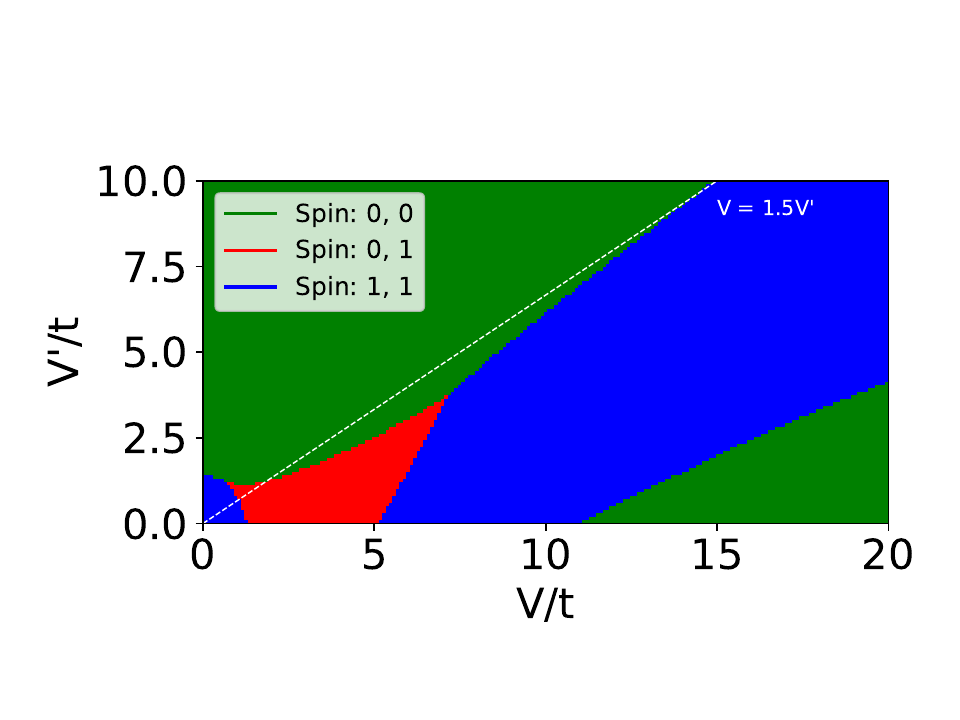}}
	\adjustbox{trim={.05\width} {.15\height} {0} {.25\height},clip}{\includegraphics[width=.99\columnwidth]{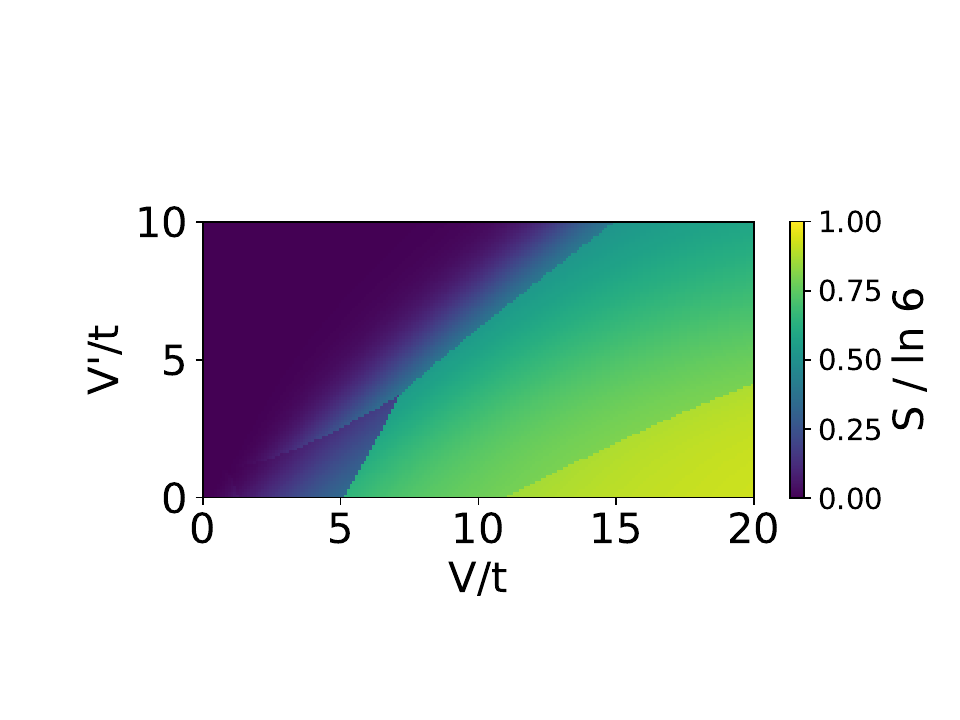}}
	\caption{{\bf Top:} Ground-state spin phase as a function of $V$ and $V'$. The line $V=1.5V'$ is shown for reference. {\bf Bottom:} Entanglement entropy between layers for the ground state.}
	\label{fig:vvp}
\end{figure}

In fig. \ref{fig:vvp}, we plot the full numerically-calculated phase diagram for all values of $V$ and $V'$, as well as the entanglement entropy of the ground state. The results presented above correspond to cuts along both axes, as well as the line $V=V'$. From this diagram we can see that for values of $V$ less than $1.5V'$, the system acts like two weakly coupled independent layers, whereas for $V>1.5V'$, the two layers become closely entangled and are strongly correlated.

\subsection{Electron Case}

For completeness, we compare the above results with the cases where one or both layers of the quantum dot plaquette are occupied by electrons rather than by holes. We find that if electrons are present rather than holes, the complex phase transitions we discussed above vanish.

\begin{figure}[!htb]
	\includegraphics[width=.49\columnwidth]{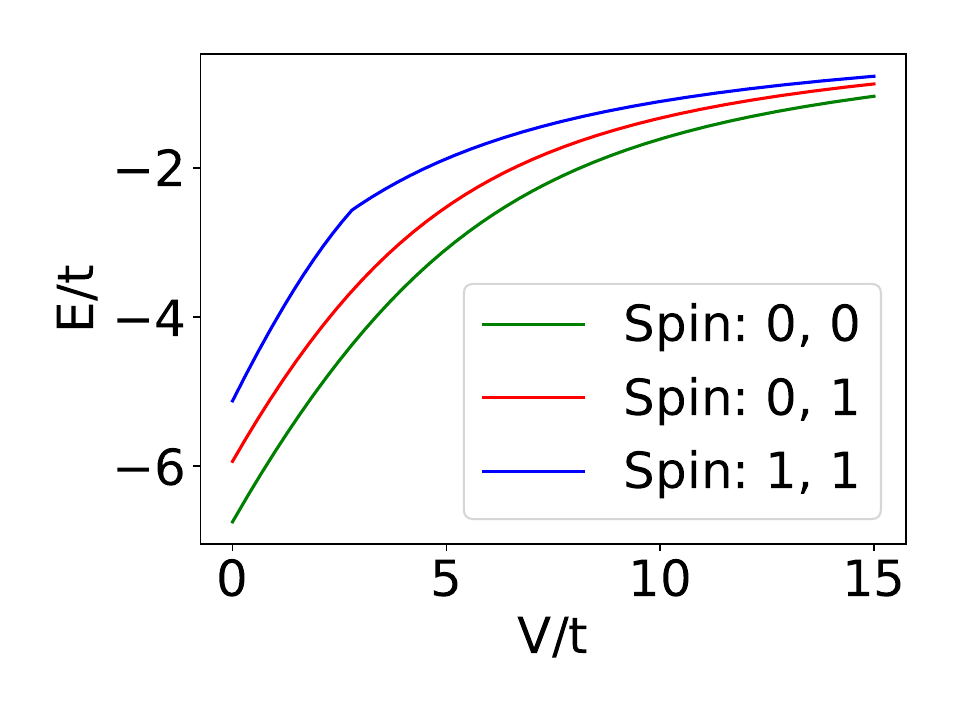}
	\includegraphics[width=.49\columnwidth]{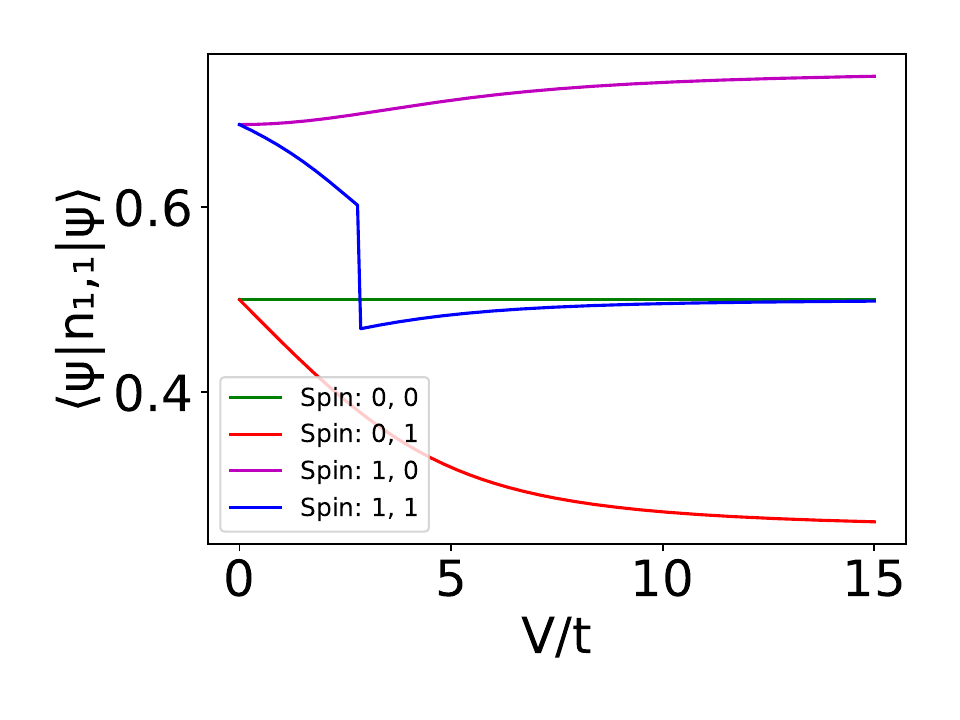}
	\adjustbox{trim={.05\width} {.15\height} {0} {.25\height},clip}{\includegraphics[width=.99\columnwidth]{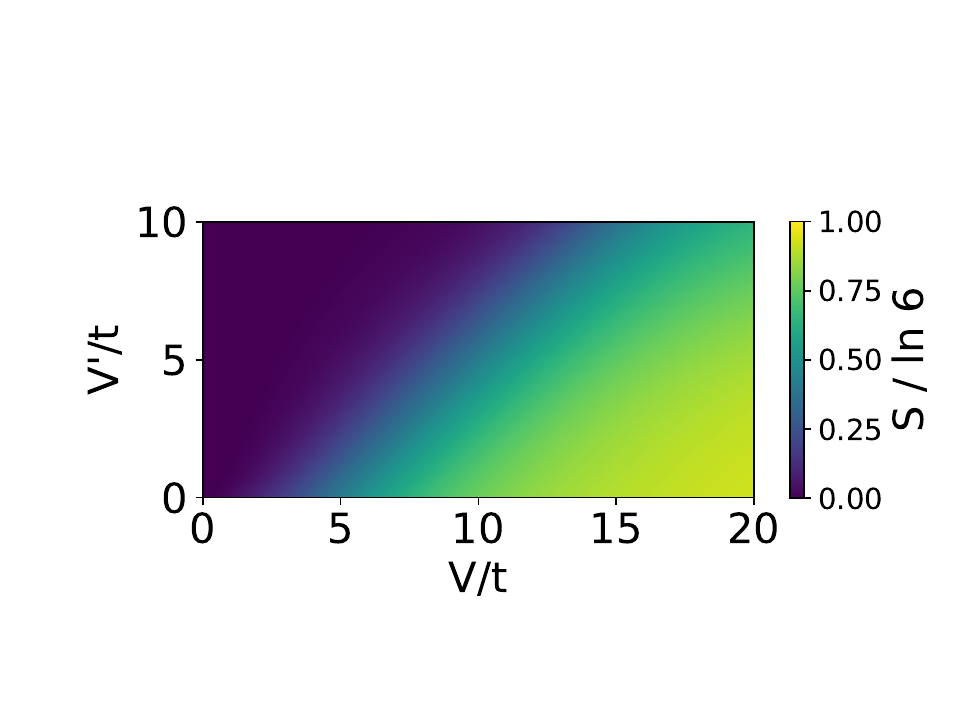}}
	\caption{{\bf Top Left:} Ground state energies versus $V/t$ for each spin configuration for a plaquette where each layer is occupied by two electrons and $V'=0$. {\bf Top Right:} $\bra{\Psi}n_{1,1}\ket{\Psi}$ for dot 1 of layer 1. {\bf Bottom:} Entanglement entropy between layers.}
	\label{fig:ee}
\end{figure}

In Fig. \ref{fig:ee}, we show the ground state energies for a bilayer plaquette with $V'=0$, and where each layer is occupied by two electrons. Here we find that the (0, 0) spin configuration is the ground state for all values of $V/t$. In fact, if $V'$ is allowed to be nonzero and vary independently of $V$, the (0, 0) spin configuration remains the ground state for any values of $V$ and $V'$. Thus, the system is always antiferromagnetic in sharp contrast to Hund's rule expectations. The absence of a phase transition occurs because when the plaquette is occupied by electrons, the ground state is spin 0 which has electron occupation $\langle n_{l,i}\rangle$ evenly distributed among all dots in the plaquette.

\begin{figure}[!htb]
	\includegraphics[width=.49\columnwidth]{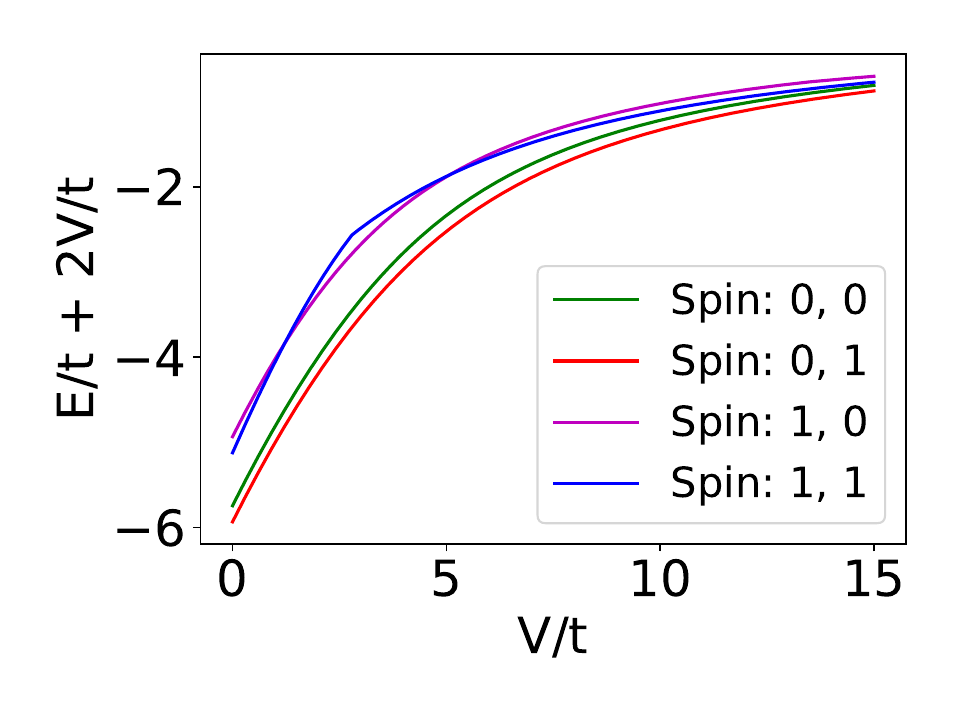}
	\includegraphics[width=.49\columnwidth]{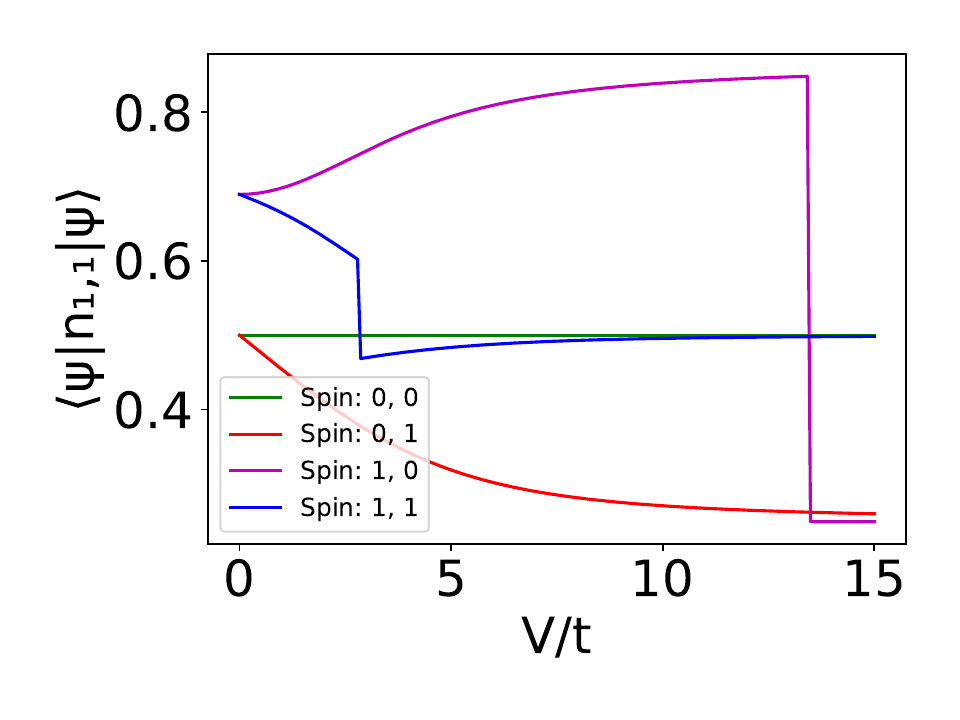}
	\adjustbox{trim={.05\width} {.1\height} {0} {.2\height},clip}{\includegraphics[width=.45\columnwidth]{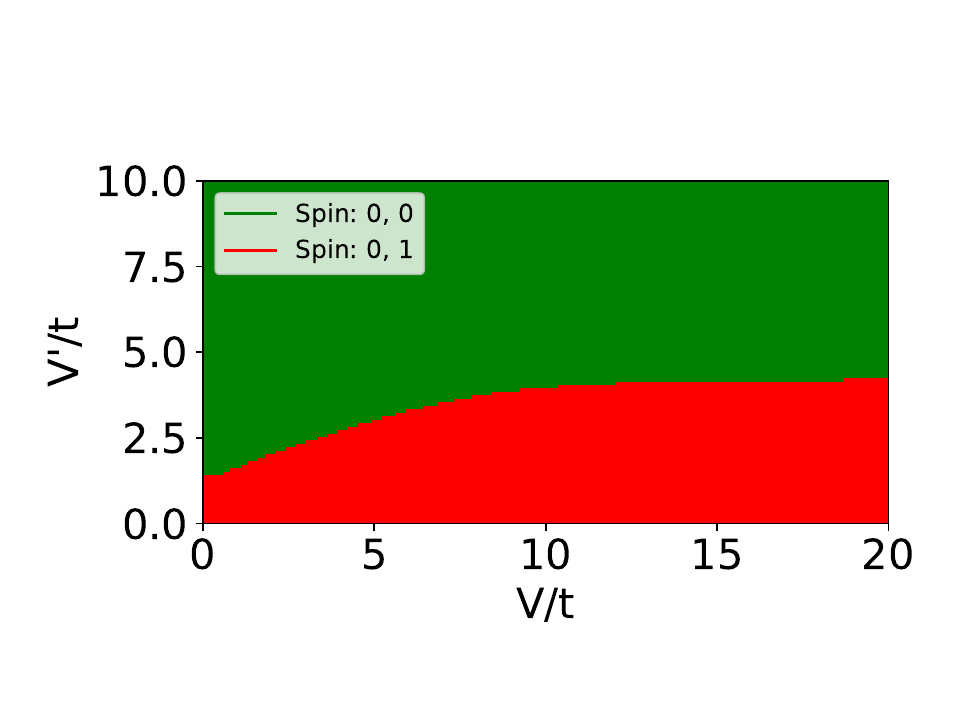}}
	\adjustbox{trim={.05\width} {.15\height} {0} {.25\height},clip}{\includegraphics[width=.55\columnwidth]{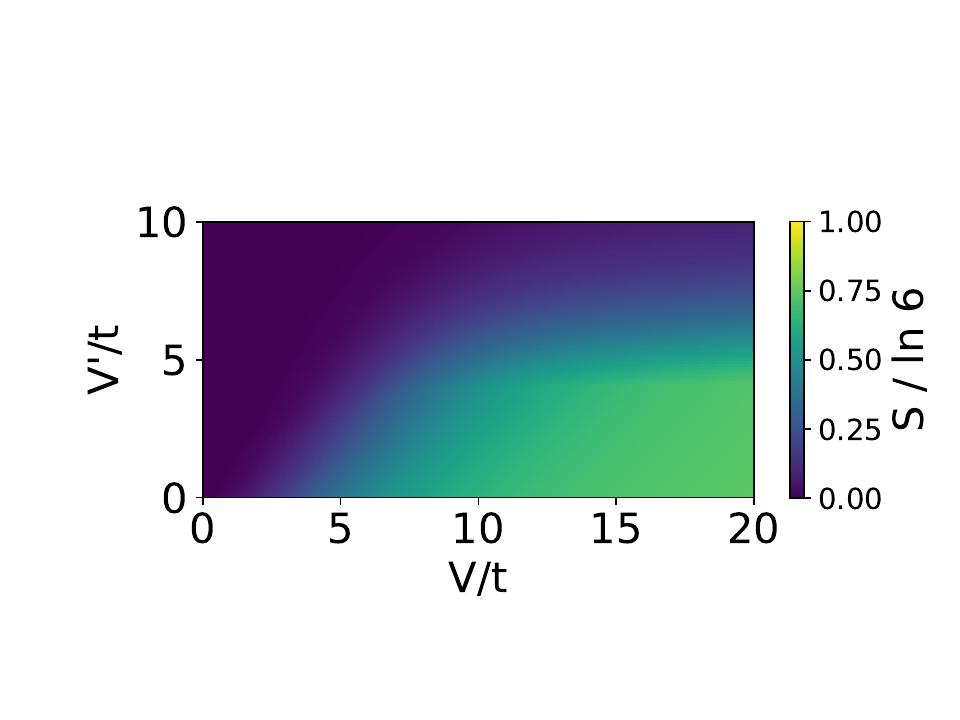}}
	\caption{{\bf Top Left:} Ground state energies versus $V/t$ for each spin configuration for a plaquette where layer 1 is occupied by two electrons and layer 2 is occupied by two holes. Here $V'=0$. {\bf Top Right:} $\bra{\Psi}n_{1,1}\ket{\Psi}$ for dot 1 of layer 1 (one of the outside dots). {\bf Bottom Left:} Phase diagram showing the spin configuration of the ground state as a function of $V$ and $V'$. {\bf Bottom Right:} Entanglement entropy between layers.}
	\label{fig:eh}
\end{figure}

It is interesting to consider the case where one layer is occupied by electrons, and the other layer is occupied by holes. In this case, the interlayer Coulomb interaction will be attractive rather than repulsive, and so in the large-$V$ limit holes in one layer will occupy the same dots as particles in the other layer. However, the electron layer's preference to equally occupy all four dots in a spin 0 state causes the spin transition of this model to remain trivial. As shown in fig. \ref{fig:eh}, there is only one spin transition, the one in $V'$ discussed prior that is present in a single layer occupied by holes.

\section{Conclusion}

We have studied the zero-temperature magnetic ground states of a bilayer quantum dot plaquette using the generalized Hubbard model with long-range intralayer and interlayer Coulomb interactions for systems of holes, electrons, and both together. A bilayer rhombus plaquette at 3/4 filling demonstrates a magnetic phase transition as the strength of the long-range Coulomb interaction is varied. A simple transition from spin 1 to spin 0 occurs in a single layer. However, interlayer effects can occur if two identical layers are brought close to one another. This is particularly noteworthy, as the spin state in one layer can directly affect the spin state in the other layer, despite the absence of interlayer tunneling or exchange interactions. This capacitive coupling is due to the holes' preference to more heavily occupy the inner two dots in a rhombus plaquette in the absence of long-range Coulomb interactions. When the Coulomb interaction strength is increased enough, the holes are pushed away from the center, which causes a phase transition. This effect is dependent on the exact plaquette geometry, as the properties of the ground state can change quite drastically for various layouts of quantum dots. In fact, for many other geometric layouts of only a few quantum dots (which we have investigated using the same techniques as presented in this work), the addition of long-range Coulomb interactions has little to no qualitative effect on the properties of the ground state. We showed that a simple magnetic phase transition occurs for a rhomboidal geometry, even in a single-layer, and that additional two-layer effects occur so long as the interlayer Coulomb interaction is greater than roughly 1.5 times the intralayer Coulomb interaction strength.

Among our interesting and nonintuitive findings are complex magnetic phase transitions in the bilayer hole system, but not necessarily in the bilayer all-electron or electron-hole systems.  Our predictions of nontrivial magnetic phases and transitions among them can be tested in existing bilayer quantum dot structures recently fabricated in Delft \cite{TidjaniARXIV2023}.  We believe such studies of quantum ground states to be most promising initial experimental investigations of hole quantum dot structures since studying quantum correlations is one of the main goals of quantum circuits being developed worldwide in different platforms.

As quantum dot experimental capabilities continue to grow, the capability to observe physical phenomena in quantum dot devices is an important step towards realizing quantum technologies. Magnetic order and effects such as the phase transitions discussed in this work are examples of highly nontrivial phenomena which are particularly well-suited for the current stage of experimental capabilities due to the small number of requisite quantum dots and the distinct qualitative features of the system's ground state. A rhombus plaquette is a good candidate for observing magnetic effects in a bilayer quantum dot system, and a simpler phase transition is present in a single-layer plaquette with the same geometry as well.

\acknowledgements

This work is supported by the Laboratory for Physical Sciences.

\bibliography{bilayerbib}

\end{document}